\documentclass[aps,prb,twocolumn,groupedaddress,amsmath,amssymb,showpacs]{revtex4}

\usepackage{graphicx}
\usepackage{hyperref}
\usepackage{pifont}

\newcommand{\homo}{\epsilon_\mathrm{HOMO}}
\newcommand{\EF}{E_\mathrm{F}}
\newcommand{\ho}{\mathrm{H}_2\mathrm{O}}
\newcommand{\Vac}{V_\mathrm{vacuum}}

\begin{document}

\title{Finite-bias electronic transport of molecules in water solution}

\author{I. Rungger$^1$, X. Chen$^1$, U. Schwingenschl{\"o}gl$^2$ and S. Sanvito$^1$}
\affiliation{$^1$ School of Physics and CRANN, Trinity College, Dublin 2, Ireland}
\affiliation{$^2$ KAUST, PSE Division, Thuwal 23955-6900, Kingdom of Saudi Arabia}

\date{\today}

\begin{abstract}
The effects of water wetting conditions on the transport properties of molecular nano-junctions are investigated theoretically by using a 
combination of classical molecular dynamics and first principles electronic transport calculations. These are at the level of the 
non-equilibrium Green's function method implemented for self-interaction corrected density functional theory. We find that water 
effectively produces electrostatic gating to the molecular junction, with a gating potential determined by the time-averaged water dipole field. 
Such a field is large for the polar benzene-dithiol molecule, resulting in a transmission spectrum shifted by about 0.6~eV with respect to that of 
the dry junction. The situation is drastically different for carbon nanotubes (CNTs). In fact, because of their hydro-phobic nature the gating is almost 
negligible, so that the average transmission spectrum of wet Au/CNT/Au junctions is essentially the same as that in dry conditions. 
This suggests that CNTs can be used as molecular interconnects also in water-wet situations, for instance as tips for scanning tunnel 
microscopy in solution or in biological sensors.

\end{abstract}


\maketitle

\section{Introduction}

Electron transfer in biological systems is rather different from electron transport in an electronic device. In addition to the intrinsic materials 
differences between the respective conducting media, namely soft molecules in biology against inorganic semiconductors in electronics, the 
electron transport in these two situations differs for the effects and the relevance of the environment. In fact, while one wants to keep
electronic devices in dry conditions to avoid electrostatic disorder, biological electron transfer is characterized by a time-evolving
local dielectric environment. 

The boundary between the two fields, electronics and biology, becomes however blurred when one looks at the nanoscale. On the one
hand, ``conventional'' nanotechnology has expanded in the biology domain with a growing number of electronic applications requiring 
operation in wet conditions. In addition to scanning tunnel microscopy in water \cite{Bruce}, which has been available for the last two 
decades, several biological sensors have been recently proposed. These for instance include cancer markers detectors \cite{Cancer} 
and protocols for DNA sequencing \cite{DNA}. On the other hand, when a solution is confined at the nanoscale, highly ordered structures
can form at room temperature \cite{h2o}. This essentially means that under such stringent confinement a biological system 
almost behaves like a solid. 

Interestingly, numerous experiments to date on electronic transport through molecules are carried out in solution \cite{bdtexpxiao,laka}. 
Still, with a few exceptions \cite{bdtwater,colin,Cao}, most of the theoretical calculations are performed in the dry, i.e. without explicitly including
water molecules in the simulations. Therefore the question on how water affects the current-voltage characteristic in a
molecular junction remains. In general one may expect that solutions made with polar molecules, such as $\ho$, may affect
significantly the transport because of the generation of local dipole fields. However, since the time scale of a transport measurement 
is far longer than the typical molecular rearrangement of a solution, one should ask what is the time-averaged dipolar field near
the molecule of interest. This, together with the degree of localization of the relevant molecular orbitals, will determine whether
or not the wetting conditions influence the junction electronic transport. 

In order to address those fundamental questions we have performed a number of combined molecular dynamics (MD) and
quantum transport simulations for molecular junctions in water. Our computational strategy is to investigate  first the
dynamics of $\ho$ and then to evaluate the electronic transport of a representative number
of configurations, i.e. for a number of MD snapshots. In particular we have considered two rather different junctions, 
both using gold electrodes but differing for the local charge arrangement of the molecule of interest. The first is made
from a polar molecule, namely benzene-dithiol (BDT), while the second includes locally charge neutral carbon nanotubes (CNTs),
in particular a (3,3) metallic and a (8,0) insulating one. 

We find that the effect of water on the transport is that of effectively gating the molecule, therefore shifting almost rigidly
its transmission spectrum. This is rather pronounced for BDT, but only tiny in the case of CNTs, and reflects the different charge distribution 
of the two classes of molecules. We associate those differences to the time-averaged dipole field of the water.
Notably all the calculations are performed with density functional theory including appropriately corrections for the electronic structure 
of water. This is an essential condition for quantitative predictions. 

The paper is organized as follows. In the next section we describe the device set-up for the calculations and briefly the computational 
tools used. Then we analyze the transport. First we look at the electronic structure of a gold capacitor with water wetting the two 
electrodes and then we consider finite bias conductance across BDT and the CNTs. Finally we present our main conclusions.

\section{Methods}
\label{sec:methods}

Classical MD calculations are performed with the {\sc namd2} package\cite{NAMD2005}. BDT is attached to the two 
gold electrodes at the Au(111) hollow site, which has been previously calculated to be the low energy bonding position for 
BDT on Au(111) \cite{cormacbdtprb}. We define our coordinate system with the $z$ axis along the (111) direction and the 
$x$-$y$ plane orthogonal to it. All the 
calculations are carried out with periodic boundary conditions in the $x$-$y$ plane. TIP water molecules are added to a $(25 \times 17.3 \times 7.3)$~\AA$^{3}$ box intercalated 
between the Au electrodes. Note that the dimension along the $z$ axis is chosen in such a way that the 
Van der Waals distance between the Au(111) surface and the water molecules is taken into account. In total there are 83  
$\ho$ molecules in the simulation box. Since our main objective is that of examining the effects of water over the 
conductance of a Au/molecule/Au device, we always fix the atomic positions of both the molecule and the electrodes. 

The interaction between the H$_2$O molecules and gold is treated at the level of a 12-6 Lennard-Jones potential
\begin{equation}
U^\mathrm{LJ}=4\epsilon\left[\left(\frac{\sigma}{r}\right)^{12}-\left(\frac{\sigma}{r}\right)^{6}\right]\:,
\end{equation}
with parameters for Au ($\epsilon =0.039$~kcal/mol and $\sigma = 2.934$~\AA) taken from the literature [\onlinecite{GoldLJ}]. 
Periodic boundary conditions are applied with a cutoff of 12 \AA~for long-range interactions. In order to maintain the size and shape 
of the cell constant, we perform simulations in the micro-canonical ensemble, with re-initialized velocities to 300~K for every 
1000 time-steps and with a time-step of 2~fs. The trajectory is recorded every 4~ps from the initial equilibration of 1~ns to a total 
simulation time of 20~ns. For the systems comprising the CNTs, Au(111)/CNT(3,3)-H$_2$O/Au(111) and Au(111)/CNT(8,0)-H$_2$O/Au(111), 
and for the reference Au(111)/H$_2$O/Au(111) capacitor, similar conditions and procedures are followed. The parameters for the 
aromatic carbon atoms are used for CNTs.

For each system we calculate the transport properties for a set of representative MD configurations taken after equilibration.  From these 
we can then estimate the fluctuations in the transmission and current over time, as well as their time-averaged values. We use the {\sc smeagol} 
\textit{ab initio} electronic transport code to calculate the zero bias transmission coefficients and the current-voltage ($I$-$V$) characteristics. 
{\sc smeagol}  combines the non-equilibrium Green's function method with density functional theory (DFT)\cite{smeagol1,smeagol2,senepaper} 
and has the pseudopotential code {\sc siesta}\cite{siesta} as its electronic structure platform.

The local density approximation (LDA) to the exchange and correlation functional is adopted throughout. The atomic self-interaction 
correction (ASIC)\cite{dasasic} scheme however is used for the water and the BDT molecule, in order to bring their ionization potentials (IPs) 
in closer agreement to experiments. This has been already proved to be a successful strategy for aligning correctly the highest 
occupied molecular orbital (HOMO) energy, $\homo$, of the molecule under consideration to the Fermi level 
($E_\mathrm{F}$) of the electrodes\cite{cormacbdtprb,cormacbdtprl}. Note that the same corrections also reproduce well the 
band-gap of many insulating oxides, after the ASIC potential is rescaled appropriately\cite{dasasic}. Such a rescaling 
in bulk crystals is attributed to charge screening, which in solids is usually stronger than in molecules. In general we use a scaling 
parameter, $\alpha$, ranging between 0 and 1, to adjust the amount of self-interaction correction included 
($\alpha=0$ corresponds to the LDA, $\alpha=1$ is the full ASIC).  Usually $\alpha$ is 1 for small molecules, it is around 0.5 for 
insulating oxides and it vanishes for metals. For this reason ASIC is never applied to Au.

For the transport calculations we use a real space mesh cutoff of 200~Ry and an electronic temperature of $300$~K. The unit cell 
includes 5 Au atomic layers on each side of the Au(111) surface, which are enough to screen charging at the Au-molecule interface. 
In order to reduce the system size the Au $5d$ shell is kept in the core, so that we consider only $6s$ orbitals in the valence. 
We use a single-$\zeta$ basis for Au $6s$, specifically optimized to give the correct work function for the Au(111) surface.\cite{magmol} 
The rest of the basis set is double-$\zeta$ for the C $s$ and $p$ orbitals, and double-$\zeta$ plus polarization for S ($s$ and $p$). For 
the $\ho$ molecules we use a double-$\zeta$ basis for both H and O. However when calculating transport through CNTs we reduce it 
to single-$\zeta$, in order to keep the size of the density matrix tractable.

The charge density is obtained by splitting the integral of the Green's function into a contribution calculated over the
complex energy plane and one along the real axis \cite{smeagol1,smeagol2}. The complex part of the integral is computed 
by using 16  energy points on the complex semi-circle, 16 points along the line parallel to the real axis and 16  poles. The integral 
over real energies necessary at finite bias is evaluated over at least 1000 points\cite{smeagol1,smeagol2} per eV.  

\section{Results and Discussion}

\subsection{Au capacitor in water}
\label{sec:watergold}

\begin{figure}
\center
\includegraphics[width=8.0cm,clip=true]{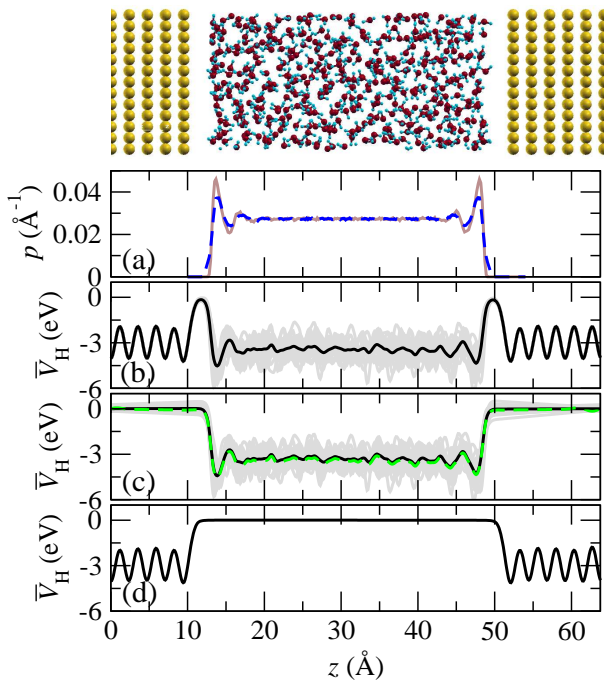}
\caption{(Color online) Au capacitor with water as dielectric medium. The top panel shows the unit cell used for the 
MD simulations, which includes $\ho$ confined between the two Au electrodes. In panel (a) we show the probability to find 
a O (solid curve) or a H (dashed curve) atom at a given $z$ position. Panels (b-d) are the planar averages 
of the Hartree electrostatic potential $\bar{V}_{\mathrm H}$ as function of position along the transport direction, $z$, for different setups: 
(b) entire junction, (c) junction without the electrodes, and (d) junction without water. The grey curves, merging in a shadow, are the 
results for 21 snapshots taken at different times after equilibration, the black solid curves are their time averages. The dashed curve 
in (c) shows the difference between the time-averages of panels (b) and (d).}
\label{Fig1a}
\end{figure}
In a transport calculation it is crucial to describe accurately both the electrodes' work function, $W$, and the IP of the molecule,
so that  the correct alignment of the molecular levels to the electrodes' $E_\mathrm{F}$ is reproduced. With this in mind we first 
analyze the electronic structure of water sandwiched between Au electrodes. Such a setup corresponds to calculating the electronic
structure of a Au parallel plate capacitor, where the two plates are separated by water. The unit cell, containing 408 $\ho$ 
molecules and 480 Au atoms, is shown in Fig. \ref{Fig1a}. The statistical distribution of $\ho$ about the Au plates
is described in Fig. \ref{Fig1a}(a), where we plot the normalized probability, $p(z)$, to find O (solid curve) or H (dashed curve) 
at a given position $z$ in the cell ($\int p(z)~dz=1$).  Such a distribution is obtained by using the MD data for all the time-steps 
included in the 16~ns simulation after equilibration. We note that $p(z)$ is constant, in the middle of the gap between the plates, indicating an 
average random arrangement of the $\ho$ molecules. In contrast close to the Au interface there are marked oscillations in 
$p(z)$, signaling a correlation of the water position with respect to the Au surface. Note that the peaks in $p(z)$ are found at the same positions for O and H atoms. This indicates that on average there is no net dipole at the interface.

Next the Au work function is calculated by using the same Au parallel plate capacitor of Fig.~\ref{Fig1a} after we have removed 
the water, i.e. with vacuum as spacer between the plates. $W$ is then the energy difference between $E_\mathrm{F}$ and the 
vacuum potential, $V_\mathrm{vacuum}$. Such an exercise is reported in Fig.~\ref{Fig1a}(d) where we show the planar average 
$\bar{V}_H$ of the electrostatic Hartree potential along the $x$-$y$ plane. Note that the absolute value of $\bar{V}_H$ in Au is 
arbitrary, since it depends on the pseudopotentials. However, in the middle of the capacitor $\bar{V}_H=V_\mathrm{vacuum}$. 
Therefore, setting $V_\mathrm{vacuum}=0$ we have $W=-E_\mathrm{F}$. We find 
$W=5.3$~eV, in good agreement with experiments.

We now turn our attention to the electronic structure of water. This must be extracted for its liquid phase, i.e. from the MD
simulations. We consider data averaged over 21 structural configurations corresponding to 
21 equally spaced MD snapshots. The so-calculated planar average of the electrostatic potential is shown in Fig. \ref{Fig1a}(b). 
The grey lines, merging in a shadow, are the superimposed curves for all the snapshots, while the black line is their time-average. 
We note oscillations of $\bar{V}_\mathrm{H}$ close to Au. These are due to the arrangement of the water molecules 
with respect to the Au surface. In contrast in the middle of the junction $\bar{V}_H$ is rather flat due to the average random 
orientation of the molecules [see also Fig.~\ref{Fig1a}(a)].

\begin{figure}
\center
\includegraphics[width=6.0cm,clip=true]{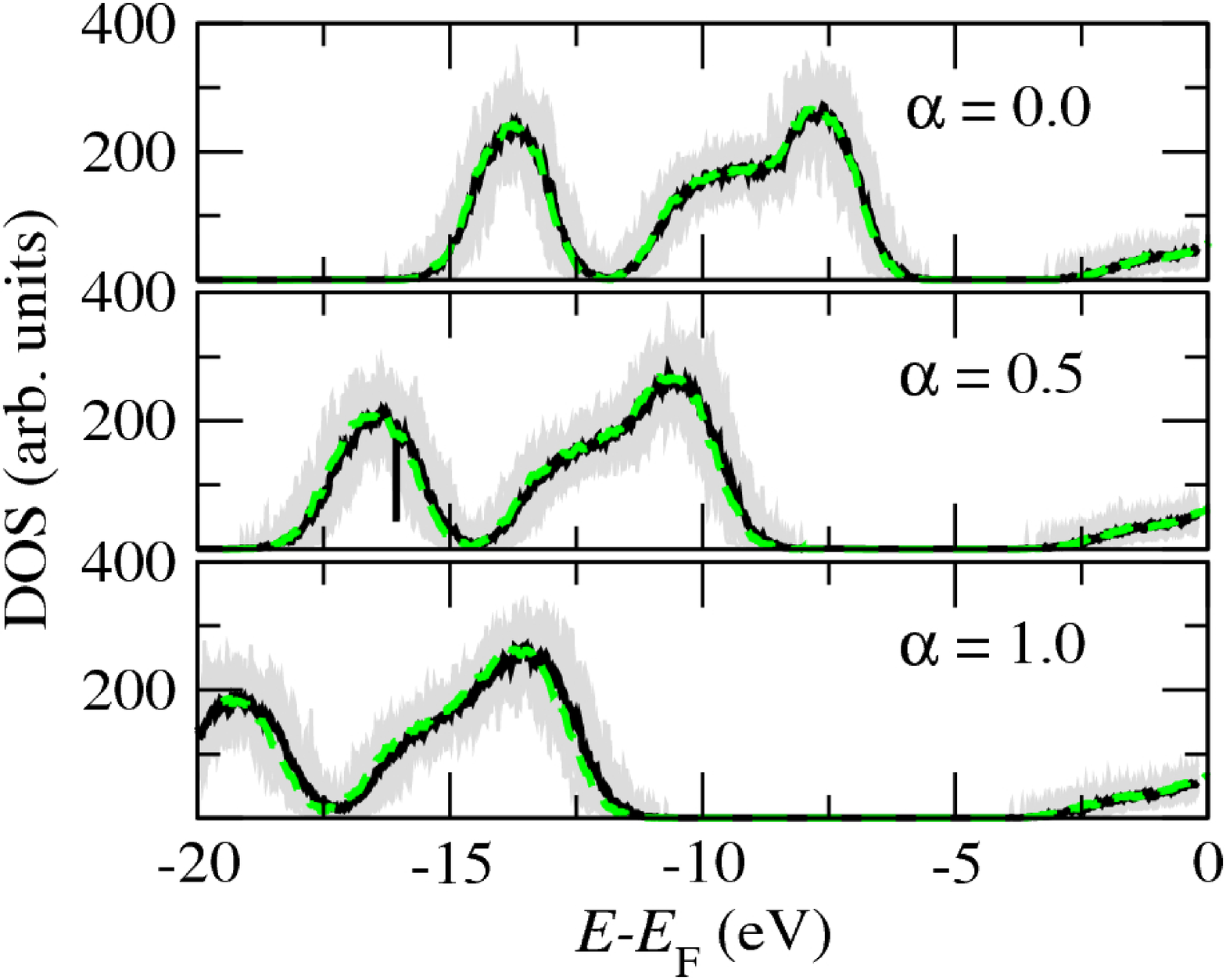}
\caption{(Color online) DOS projected onto $\ho$ for the capacitor of Fig. \ref{Fig1a} and obtained for different 
values of the ASIC scaling parameter $\alpha$. The grey lines are the superimposed curves for 21 MD snapshots and 
the solid black one is their average. The dashed curve is the average DOS for the same MD snapshots for a system where 
Au is replaced by vacuum.}
\label{Fig2a}
\end{figure}
From the same simulations we can extract the IP of liquid water. In transport one wants $\homo$ of the molecule of 
interest to correspond to the actual negative of its IP \cite{TSFB,CormacSTM}. Although this should be the case 
for exact DFT, it happens only rather rarely in practice for the standard local approximations of the exchange and correlation 
functional\cite{DD}. In the case of liquid H$_2$O the experimental value of the IP 
is in the range between 9.3 and 10~eV, in rather good agreement with recent \textit{ab initio} calculations of the electron 
removal energy (see [\onlinecite{wateriedft,wateriedft2}] and references therein). 

We evaluate $\homo$ for liquid H$_2$O by calculating the density of states (DOS) projected onto the water molecules. This is
shown in Fig.~\ref{Fig2a} ($V_\mathrm{vacuum}=0$), where again the grey curves are the superimposed data for 
all the MD snapshots and the black one is their average. We note that there are large 
fluctuations in the DOS over time with energy shifts of the order of 1~eV. The average value however is smooth and can be
reliably taken as the liquid water DOS. In this way the top of the water ``valence band'' is calculated to be only -6~eV in 
LDA, in agreement with previous DFT calculations\cite{waterdft}, but still far from the experimental value. 

We then apply ASIC to $\ho$ and find that $\homo$ moves to -8.5~eV for $\alpha=0.5$ and to -11~eV for $\alpha=1.0$, so that
$\alpha=0.7$ fits the average experimental value (-9.5~eV). Such an optimal value, as usually with ASIC\cite{dasasic}, 
in general improves the entire electronic structure and returns an HOMO-LUMO gap of about 6.4~eV, in good agreement with 
the experimental value of 6.9 eV.\cite{waterexpgap} Note that $\alpha=0.7$ is typical for moderately ionic insulating oxides\cite{dasasic}. 

In order to analyze the effects of the Au/H$_2$O interface over the water IP and DOS we perform a second set of calculations, where 
we remove the Au plates and we replace them by vacuum. This corresponds to an hypothetical H$_2$O slab. In this case [see 
Fig.~\ref{Fig1a}(c)] $\bar{V}_\mathrm{H}$ at a given MD time-step has a finite slope in the vacuum region, which is caused by 
the non-compensated dipoles at the H$_2$O external surface. These dipoles produce a long-range electric field outside slab, 
so that $V_\mathrm{vacuum}$ of a single snapshot is not defined. However the time-averaged $\bar{V}_\mathrm{H}$ (black curve) 
is approximately flat in the vacuum, demonstrating that, although at each time-step surface charge may lead to long-ranged 
electric fields, its time-average is actually zero. 

If we now take the average $\bar{V}_\mathrm{H}$ away from the water molecules as $\Vac$, we can plot the time-averaged 
DOS for the water slab and superimpose it to that calculated for the Au/H$_2$O/Au capacitor [see Fig.~\ref{Fig2a}]. We find that the
two DOSs overlap on each other, confirming our results for the water IP and the fact that on average there is little electronic interaction
between Au and H$_2$O. Our results also suggest that one should ideally use periodic boundary conditions to simulate the electronic 
structure of molecules in solution. These eliminate the possible spurious electric fields in the vacuum, which can 
lead to an unphysical rearrangement of the energy levels. Furthermore for simulations of $\ho$ surfaces it is essential to consider 
time-averages, so that the water electric field vanishes in vacuum. 
Fig.\ref{Fig1a}(c) also shows the difference between the time-averaged $\bar{V}_\mathrm{H}$ of the Au capacitors with 
and without water (dashed line). This difference is almost identical to the time-averaged $\bar{V}_\mathrm{H}$ for the water slab
and is consistent with the fact that the Au electrodes do not induce any noticeable change in the average DOS of H$_2$O.

\section{Benzene-dithiol}
\label{sec:BDT}

\begin{figure}
\center
\includegraphics[width=5.5cm,clip=true]{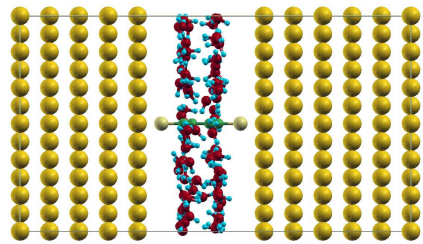}
\caption{(Color online) Unit cell used for the BDT molecule in water attached to Au electrodes.}
\label{Fig3a}
\end{figure}
Electron transport through BDT attached to Au electrodes has been extensively studied both 
experimentally\cite{bdtexpreed,bdtexpxiao,bdtexp2,bdtexp3} and theoretically.\cite{cormacbdtprl,cormacbdtprb} In fact, because of its simple 
structure, BDT is an ideal system for comparing theory with experiments. However, also for BDT the LDA description is not 
adequate and it is necessary to employ ASIC\cite{cormacbdtprl,cormacbdtprb}. In order to limit the number of adjustable 
parameters we set the same $\alpha$ for both BDT and H$_2$O, and check that such a value reproduces well previous transport 
calculations for the Au/BDT/Au junction in dry conditions\cite{cormacbdtprl,cormacbdtprb}. 

\begin{figure}
\center
\includegraphics[width=6.0cm,clip=true]{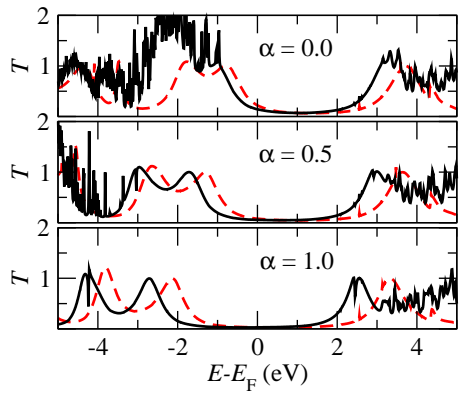}
\caption{(Color online) Transmission coefficient for transport through BDT in water as function of energy for one MD snapshot 
and for different values of the ASIC scaling parameter $\alpha$. The dashed curves are for BDT in dry conditions 
(no water), while the solid curves are for BDT in solution.}
\label{Fig4a}
\end{figure}

We first investigate the $V=0$ transport (the cell used is shown in Fig. \ref{Fig3a}). In Fig. \ref{Fig4a} the transmission 
coefficient $T(E;V=0)$ for one MD snapshot is shown as function of energy, $E$, for different values of the ASIC scaling parameter 
$\alpha$ (solid curves). The same quantity is compared to that calculated for the same cell, this time without including water (dashed curves). 
In general the effects of water are two-fold: firstly there is a shift of the BDT transmission peaks to lower energies, and secondly there appear 
additional sharp transmission peaks, which are attributed to resonant transport through the electronic states of H$_2$O. Interestingly the 
height, the width and the relative position of the transport peaks with respect to each other is unchanged when water is present. 

Therefore the main effect of adding water is to shift the energy levels of BDT, so that water acts as an external gate. Since the BDT molecular
orbitals extend over the entire molecule and are strongly coupled to Au\cite{cormacbdtprb}, all the levels shift by approximately the 
same amount. If the levels were more localized, we might have expected a change in their relative position, sensitively dependent 
on the local configuration of $\ho$ \cite{LocField}.

\begin{figure}
\center
\includegraphics[width=6.0cm,clip=true]{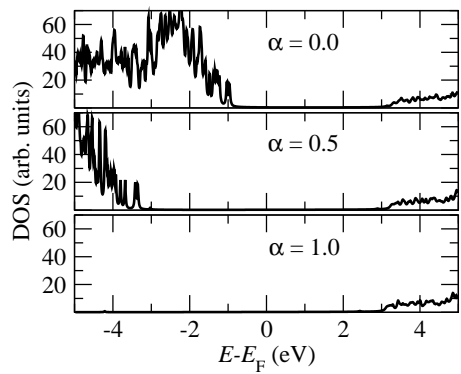}
\caption{Density of states projected onto $\ho$ for the junction of Fig. \ref{Fig3a} as function of energy for one MD snapshot 
and for different values of the ASIC scaling parameter $\alpha$.}
\label{Fig5a}
\end{figure}

We now analyze in more detail the electronic properties of H$_2$O for the same MD snapshot. In Fig. \ref{Fig5a} the DOS projected 
onto the water molecules is shown over the same energy range as that of $T$ in Fig. \ref{Fig4a}. It is clear that the additional 
peaks in the transmission (Fig.~\ref{Fig4a}) are at energies where H$_2$O has a finite DOS, confirming that these are due to transport 
through the electronic states of water. The additional transmission peaks below $E_\mathrm{F}$ are rather close to the Fermi level
when $\alpha=0$, whereas they move down in energy as $\alpha$ is increased. In contrast the position of the peaks above 
$E_\mathrm{F}$ is almost constant for different $\alpha$, since the ASIC mainly affects occupied states. We also find that 
for $\alpha=0$ the water HOMO is about -6.3~eV from $V_\mathrm{vacuum}$, while it is at -8.7~eV for $\alpha=0.5$. 
These values agree with the values obtained for the water slab.

\begin{figure}
\center
\includegraphics[width=6.0cm,clip=true]{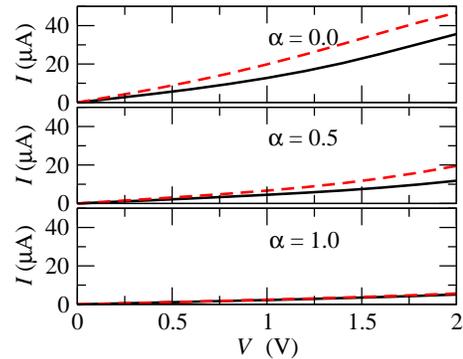}
\caption{(Color online) Current, $I$, as function of the bias voltage, $V$, through the BDT molecule calculated at the first MD 
time-step and for different values of the SIC scaling parameter $\alpha$. The dashed curves are for the molecule in the dry
while the solid curves are for BDT in solution.}
\label{Fig6a}
\end{figure}

The self-consistent current-voltage, $I$-$V$, curve is shown next in Fig. \ref{Fig6a} for the same MD snapshot and for different values of $\alpha$. 
Generally speaking the presence of water leads to a reduction of the current, which is more pronounced for small $\alpha$. The reason for
such a reduction is that $\homo$ is very close to $E_\mathrm{F}$ for small $\alpha$, so that a tiny shift of the BDT levels to lower energies 
considerably reduces $T(E_\mathrm{F};V\approx0)$. For $\alpha=1$ there is almost no change in the current, since $E_\mathrm{F}$ 
is approximately at mid-gap already in dry conditions.

We now briefly discuss the most appropriate value of $\alpha$ in this context. For BDT in the gas phase $\alpha=1$ gives 
$\homo$ close to the experimental -IP.\cite{dasasic,cormacbdtprl} However, when BDT is immersed in water additional 
screening lowers down the value of $\alpha$. We then take $\alpha=0.7$ (the optimal value for $\ho$), which provides 
a good IP for water and also accounts for the additional screening in BDT due to the solution. Importantly our results 
depend little on the exact choice of $\alpha$, as long as it is of the order of 0.5, i.e. such that $\homo$ for H$_2$O is well 
below the Au $E_\mathrm{F}$. 
Note that using $\alpha=0$ would lead to the erroneous prediction that water becomes conducting at about 1 V of bias.

\begin{figure}
\center
\includegraphics[width=7.0cm,clip=true]{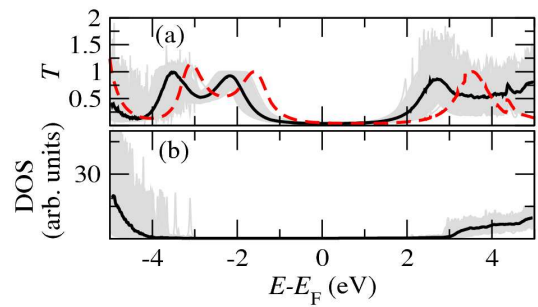}
\caption{(Color online) Time averaged (a) transmission coefficient and (b) DOS projected onto the $\ho$ molecules for BDT attached to Au. 
The calculations are obtained with ASIC and $\alpha=0.7$; the solid black curves are the time-averages over 201 
time-steps while the grey ones forming a shadow are for each of the 201 time-steps. The dashed curve is for the dry situation.}
\label{Fig7a}
\end{figure}

We now move to calculate the time-averaged transmission coefficient and the $I$-$V$ curves. In this case $T(E; V=0)$ and the $\ho$
DOS are evaluated over 201 snapshots taken in the last 16~ns of our MD simulations, while the $I$-$V$ curves are evaluated over 
only 21. Transmission and DOS are presented in Fig.~\ref{Fig7a}, where again the curves for the single snapshot calculations are 
plotted in grey to form a shadow, while their average is a solid black line. In general, when $\ho$ is introduced in the simulation,
there is a rigid shift of the entire spectrum towards lower energies with respect to the dry situation (dashed line).
This is because BDT and $\ho$ are both polar molecules and in time the water molecules arrange around BDT so to screen the 
local dipole field. Such screening moves the average BDT molecular levels to lower energies. 

\begin{figure}
\center
\includegraphics[width=7.0cm,clip=true]{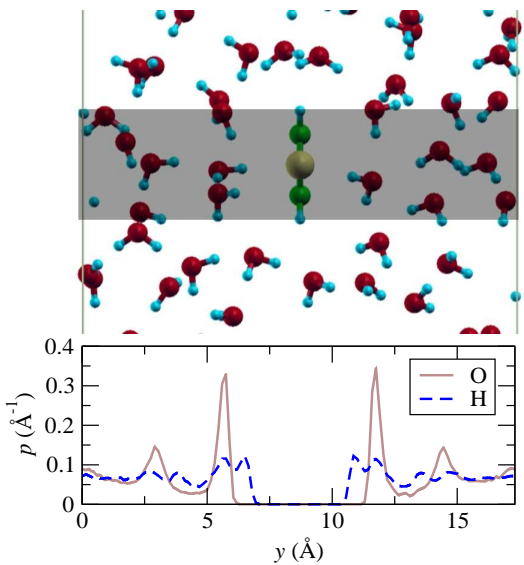}
\caption{(Color online) Probability to find an O (solid curve) or a H (dashed curve) atom at a given position $y$ in the Au/BDT/Au junction, for atoms whose $x$ coordinate lies within the shadowed region. In the top panel we show a representative snapshot of the atomic configuration of the water.
Note that in the first solvation layer the $\ho$ molecules are oriented with the H atoms pointing towards the BDT.}
\label{Fig8a}
\end{figure}

This analysis is confirmed in Fig.~\ref{Fig8a}, where we present the O and H position distributions along the $y$ direction, for those H and O atoms lying either above or below the plane of the BDT (shadowed region in Fig. \ref{Fig8a}). Note that we define the $y$ axis as the direction perpendicular to the plane of the BDT. In contrast to the 
case of the Au capacitor, now the $p(y)$'s of O and H ions differ near BDT. In particular we find that H approaches
BDT approximately 1~\AA\ closer than O. This means that on average the first solvation layer is oriented with the H atoms 
of $\ho$ molecules pointing towards the BDT, as suggested by elementary electrostatics, since the C and S atoms are fractionally negatively charged, while the H atoms have a positive charge. The second peak of the H atoms overlaps with the first peak of O, indicating that while one of the H atoms points towards the BDT, the second one aligns with the negative O atoms. It is also interesting to note that
the O distribution has a second pronounced peak in addition to that close to the BDT, signaling a relative large degree of order
also in the second solvation layer. 

We finally turn our attention to the fluctuations. Generally, time-fluctuations in the position of the BDT single particle levels result in 
zero-bias conductance fluctuations\cite{bdtwater}. These cause both a reduction in the average height of the various transmission peaks and at the 
same time an increase of their widths. For BDT attached to Au this second effect is rather small, since the transmission peaks at each time-step 
are already rather broad, due to the strong electronic coupling with the electrodes. However we expect that for molecules weakly coupled 
to the electrodes and thus presenting sharp peaks in $T(E)$ this effect will be more pronounced, probably dominating the energy level
broadening.

\begin{figure}
\center
\includegraphics[width=6.0cm,clip=true]{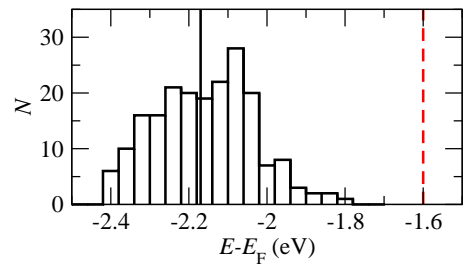}
\caption{(Color online) Histogram of $\homo$ extracted from the maximum in $T(E)$ at around -2~eV (see Fig.~\ref{Fig7a}) for $\alpha=0.7$ 
and for 201 MD time-steps. $N$ is the number of times $\homo$ is found in the given energy window specified by the bin width. The solid 
line indicates the time-averaged $\homo$, while the dashed red one marks the result in dry conditions.}
\label{Fig9a}
\end{figure}

As already mentioned before the electrostatic interactions of water with the BDT mimics a gate potential. 
At each MD time-step such an effective gate voltage changes, depending on the relative position of $\ho$. 
In order to quantify the fluctuations of the BDT molecular levels we track the position of $\homo$ [from the peak in $T(E)$] 
over time, and display the result in the form of a histogram in Fig.~\ref{Fig9a}. In the plot $N$ is the number of counts $\homo$
is found in a particular energy window, the dashed red line indicates the position of $\homo$ in the dry, while the solid 
black line indicates the time averaged $\homo$ in $\ho$ solutions. Clearly $\homo$ fluctuates between -1.8~eV and -2.4~eV, 
i.e. in an energy range of 0.6~eV. The time-averaged $\homo$ is about 0.6 eV below $\homo$ for the dry molecule, which means that the effective water-induced gating potential is about 0.6 eV.
\begin{figure}
\center
\includegraphics[width=6.0cm,clip=true]{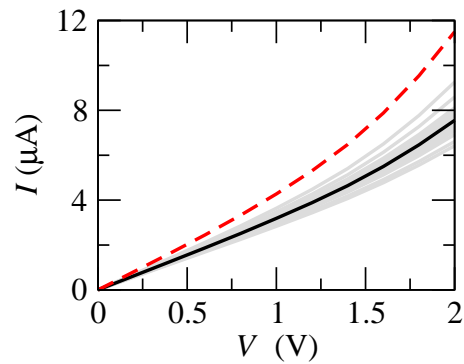}
\caption{(Color online) $I$-$V$ curve for a BDT molecule attached to gold in presence of water obtained with ASIC ($\alpha=0.7$); the solid black 
curve corresponds to the average current over 21 time-steps, the grey curves merging in a shadow are the $I$-$V$ curves of each individual configuration and the red dashed curve is for BDT in dry conditions.}
\label{Fig10a}
\end{figure}
Finally in Fig.~\ref{Fig10a} we present the self-consistent $I$-$V$ characteristics, where we conclude that the current fluctuates about its time average 
by approximately $\pm20\%$, while it is reduced from that in the dry by about 35\%. This discussion is based on ASIC calculations with the optimal 
value of $\alpha=0.7$. As discussed previously, the most appropriate correction for BDT in the dry is $\alpha\approx1$. If the same correction is
exported into the wet situation the reduction of the current due to water becomes almost negligible. 

\section{Carbon nanotubes}
\label{sec:cnt}

We now move to the analysis of the effects of $\ho$-wetting on non-polar molecules, i.e. molecules presenting local charge neutrality. In particular we choose 
two different CNTs: i) (3,3) metallic armchair and ii) (8,0) insulating zigzag.

\begin{figure}
\center
\includegraphics[width=8.0cm,clip=true]{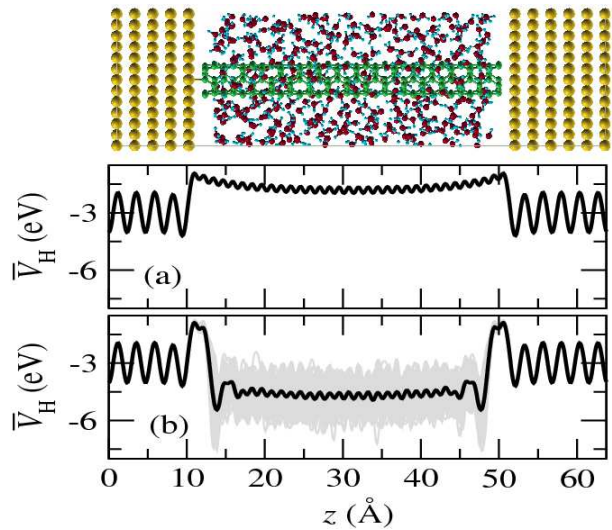}
\caption{(Color online) Au/CNT/Au junction. The top panel shows the unit cell used for the MD simulations, which includes the CNT, Au electrodes and
water molecules. In panels (a) and (b) we present the planar averages of the Hartree electrostatic potential, $\bar{V}_\mathrm{H}$, as function of position 
along the transport direction $z$: (a) junction without water, and (b) junction with water. The grey curves, merging in a shadow, are the results for all 201 
MD snapshots, the black solid curves are their time-averages.}
\label{Fig11a}
\end{figure}

In the case of the metallic (3,3) CNT we again perform 20~ns long MD simulations and calculate the observables over 201 equally spaced snapshots in the 
last 16~ns. The unit cell used is shown in Fig.~\ref{Fig11a} for one particular MD snapshot. This has a (20.0$\times$17.3)~\AA$^2$ cross section and
contains 480 Au atoms, 192 C atoms and 360 $\ho$ molecules. The Au-CNT distance is simply obtained by adding the Au and C atomic radii 
(respectively 1.44~\AA\ and 0.7~\AA) and it is close to that obtained by total energy minimization \cite{cnt33fp}. We note that the exact conformation of 
the Au-CNT bonding is not known and that changes in bond structure lead to quantitative changes in the transmission spectra.\cite{cntautightbinding} 
Here however we are mainly interested in investigating how the transmission is affected by the water so that the precise bonding 
geometry is less important. 

As already mentioned before, because of the large system size here we use a single-$\zeta$ basis for $\ho$. We verified that this gives a similar IP 
to that obtained with the double-$\zeta$ basis. In what follows we will use $\alpha=0.7$ for $\ho$, but no ASIC for the CNTs, since their
electron screening is good. We have verified that the band-structure of the (3,3) CNT agrees well with previous calculations.\cite{cnt33bands} 
In particular we obtain a CNT work function of 4.4~eV in good agreement with previous calculations\cite{calcswc33}. The IP for (3,3) CNT is not
available experimentally, but that of similar CNTs ranges between 4.8~eV and 5.0~eV,\cite{expwc33,expwc33p2,expwc33p3} thus is not 
far from what calculated here. 

Since $W$ of Au is about 1~eV larger than that of the CNT, electrons transfer from the CNT into Au, leading to a substantial band-bending. 
This is demonstrated in Fig.~\ref{Fig11a}(a), where the planar average of the Hartree potential is plotted for the Au/CNT/Au junction in dry conditions. 
Close to the Au/CNT interface the oscillating $\bar{V}_\mathrm{H}$ is higher than in the middle of the junction, whereas for an infinite CNT $\bar{V}_\mathrm{H}$ oscillates around a constant average.
We note that such charging effects have been neglected in previous tight-binding calculations.\cite{cntautightbinding} 
Importantly however charging leads to a shift in the transmission spectrum, so that it is important to include such an effect in a self-consistent way.

Next we look at the wet situation of Fig.~\ref{Fig11a}(b). In this case $\bar{V}_\mathrm{H}$ for a single MD snapshot oscillates dramatically along the CNT. 
However, when the time average is considered [solid black curve in Fig. \ref{Fig11a}(b)] a regular pattern emerges, where $\bar{V}_\mathrm{H}$
resembles closely that obtained in the dry. This confirms that on average the position of the $\ho$ molecules away from the interface is random. 
Since CNTs are hydro-phobic, we expect the interaction between the water and the CNT to be weak. This is confirmed by taking the difference between 
$\bar{V}_\mathrm{H}$ calculated with and without $\ho$ molecule and observing that the resulting curve matches closely that of the water slab 
calculated previously [see Fig.~\ref{Fig1a}(c)]. 

\begin{figure}
\center
\includegraphics[width=6.0cm,clip=true]{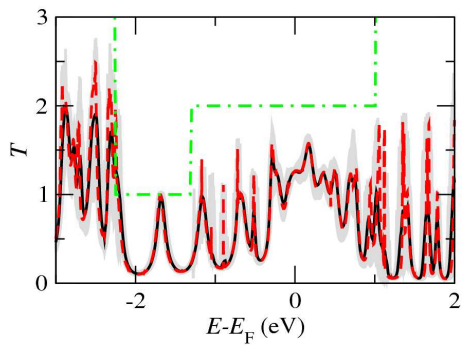}
\caption{(Color online) Transmission coefficient as function of energy for the Au(111)/CNT(3,3)-H$_2$O/Au(111) junction. The solid black curve corresponds 
to the average $T$ over 201 MD snapshots, the dashed red curve is the transmission for the same junction in the dry and the dot-dashed green 
line indicates the number of channels (spin-degenerate) for the infinite CNT. The grey curves, merging in a shadow, are $T(E;V=0)$ for each of the MD snapshot.}
\label{Fig12a}
\end{figure}
The transmission coefficients for all the 201 MD snapshots are shown in Fig. \ref{Fig12a} as super-imposed grey curves together with their average (solid black line),
the same quantity calculated in dry conditions (dashed red line) and the total number of open channel in the CNT (dot-dashed green line). In the figure we shift
$\EF$ in such a way that $\bar{V}_\mathrm{H}$ for the infinite CNT matches $\bar{V}_\mathrm{H}$ of the CNT attached to Au without water in the middle of the 
junction. The necessary shift is about 0.8~eV, which correctly corresponds to the difference in the work functions between the CNT and Au.

The main result is that the average transmission in wet conditions and that of the dry junction overlap almost exactly, demonstrating that in this case $\ho$ has 
no gating effect. This can be easily understood by recalling that, since the CNT has no polar edges, the average $\ho$ conformation presents no net electrical 
dipole, so that on average there is no shift of the CNT energy levels. Of course, each individual MD snapshot displays a dipole and the CNT energy levels get shifted. 
This leads to fluctuations in the transmission. As a result of the dipole fluctuations, we find that that some of the sharp transmission peaks visible in the dry are 
broadened up and have an average reduced height in solution. In some extreme cases (see for instance the sharp peak at  about -1~eV) they are
completely washed out by the fluctuations. 

\begin{figure}
\center
\includegraphics[width=6.0cm,clip=true]{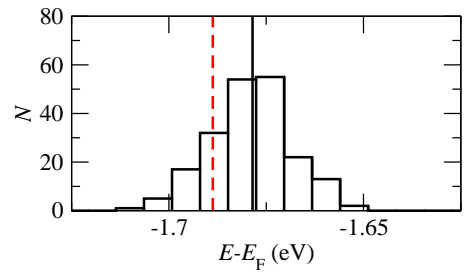}
\caption {(Color online) Histogram of the energy position of the transmission peak located at -1.7~eV below $\EF$ (see Fig.~\ref{Fig12a}). 
The histogram has been constructed from 201 MD snapshots. $N$ is the number of times the peak is found in the given energy window specified 
by the bin width. The solid line indicates the time-averaged position, while the dashed red one marks the result in dry conditions.}
\label{Fig13a}
\end{figure}

Again in order to quantify the fluctuations of $T(E)$, we choose a particular molecular level (transmission peak) and follow its 
time fluctuations. Here we select the well-defined peak at -1.7~eV below $\EF$ and present its energy distribution histogram 
in Fig.~\ref{Fig13a}. This time the peak position fluctuates over the tiny energy range of 0.06~eV, which is one order of magnitude smaller than 
that of the HOMO of BDT (see Fig. \ref{Fig9a}). The origin of such small fluctuations is twofold: firstly the interaction between $\ho$ and the CNT 
is very weak due to the hydro-phobic nature of the nanotube and secondly the CNT $\pi$-like molecular states are delocalized, so that local fluctuations 
in the electrostatic potential largely cancel out over the entire molecule. We also find that the difference between the average peak position (solid 
black line in Fig. \ref{Fig13a}) and that in the dry (dashed red line) is only 0.01 eV. This is also much smaller then the same quantity calculated
for BDT (0.6~eV). 

\begin{figure}
\center
\includegraphics[width=6.0cm,clip=true]{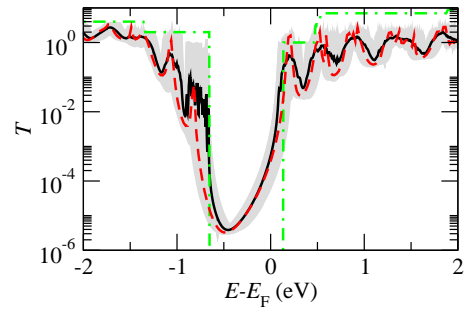}
\caption{(Color online) Transmission coefficient as function of energy for the Au(111)/CNT(8,0)-H$_2$O/Au(111) junction. The solid (black) curve 
corresponds to the average $T$ over 41 MD snapshots, the dashed (red) curve is the transmission for the same junction in the dry and the 
dot-dashed (green) line indicates the number of channels (spin-degenerate) for the infinite CNT. The grey curves, merging in a shadow, are 
$T(E;V=0)$ for each of the MD snapshot.}
\label{Fig14a}
\end{figure}

Finally we discuss results for the insulating (8,0) CNT. The simulation cell is identical to that of the (3,3) case, but this time we have 288 C atoms 
and 322 $\ho$ molecules. The MD simulations are run for 20~ns and only 41 snapshots are taken within the last 16~ns. We have reduced the number of
snapshots from 201 to 41 because this time we do not perform a detailed statistical analysis of the peak position. Our main results are shown in Fig.~\ref{Fig14a}, 
where we present $T(E;V=0)$ for all the snapshots (grey lines), their average (solid black line), that in the dry conditions (dashed red line) together
with the total number of scattering channels. The transmission coefficient is plotted in logarithmic scale in order to emphasize
the tunneling behavior in the gap. In general the quantitative features of the Au/CNT(8,0)-H$_2$O/Au junction are similar to those of the
Au/CNT(3,3)-H$_2$O/Au one. In particular also here the average transmission almost overlaps with that of the dry junction meaning that there 
is a negligible average gating. 

However for energies corresponding to the CNT gap, the transmission fluctuations are rather large due to the tunneling nature of the transport. 
For instance $T(\EF;V=0)$ fluctuates between $6.0~10^{-5}$ and $2.0~10^{-3}$ within the 41 MD snapshots considered. This means that
in tunneling conditions the presence of $\ho$ produces substantial variations in the instantaneous conductance amplitude. However and most 
importantly the time-averaged transmission at $\EF$ is only about 30\% larger than that of the dry junction. This gives us the important result that 
in general the transport in CNTs is little affected by $\ho$ solution regardless of the metallic state of the CNT. 

\section{Conclusions}

In conclusion, we have investigated the effects of water on the transport properties of two types of molecules. This is done by combining classical 
molecular dynamics with \textit{ab initio} electron transport calculations. Firstly, as an important technical result, we find that self-interaction
corrections are fundamental for describing the $\ho$ ionization energy and its band-gap. This is a pre-requisite for quantitative transport calculations.
Then our main result is the finding that the $\ho$-wetting conditions effectively produce electrostatic gating to the molecular junction, with a gating potential 
determined by the time-averaged water dipole field. Such a field is rather large for the polar BDT molecule, resulting in an average transmission
spectrum shifted by about 0.6~eV with respect to that of the dry junction. In contrast, the hydro-phobic nature of the CNTs leads to almost negligible 
gating, so that the average transmission spectrum for Au/CNT/Au is essentially the same as that in dry conditions, regardless of the CNT metallic state. 
This suggests that CNTs can be used as molecular interconnects also in water wet situations, for instance as tips for scanning tunnel microscopy in 
solutions or in biological sensors. 

\section{Acknowledgments}

This work is sponsored by Science Foundation of Ireland (grants 07/RFP/PHYF235 and 07/IN.1/I945) and by the EU FP7 (NANODNA). Computational resources have been provided by KAUST.  We thank C.D.~Pemmaraju for useful discussions.

\end{document}